\documentstyle[12pt,epsfig]{article}

\def \sect #1 {\setcounter{equation} 0\section{#1}}
\def \be  {\begin{equation}}
\def \ee  {\end{equation}}
\def \ba  {\begin{eqnarray}}
\def \ea  {\end{eqnarray}}
\def \baa {\begin{eqnarray*}}
\def \eaa {\end{eqnarray*}}
\def \bb  {\begin{thebibliography}}
\def \eb  {\end{thebibliography}}

\def \lab #1 {\label{#1}}

\def \fracs #1#2 {\mbox{\small $\frac{#1}{#2}$}}

\def \bin #1#2 {{\left({#1}\atop{#2}\right)}}
\def \as {\relax\ifmmode\alpha_s\else{$\alpha_s${ }}\fi}

\def \al #1 {\frac {\as({#1})}{\pi} }
\def \ds #1 {\ooalign{$\hfil/\hfil$\crcr$#1$}}

\def\np#1#2#3  {{Nucl. Phys.~{\bf #1} (19#3) #2}}
\def\nc#1#2#3  {{Nuovo. Cim.~{\bf #1} (19#3) #2}}
\def\pl#1#2#3  {{Phys. Lett.~{\bf #1} (19#3) #2}}
\def\pr#1#2#3  {{Phys. Rev.~{\bf #1} (19#3) #2}}
\def\prd2#1#2  {{Phys. Rev.~{\bf #1} (2000) #2}}
\def\prl#1#2#3  {{Phys. Rev. Lett.~{\bf #1} (19#3) #2}}
\def\prep#1#2#3 {{Phys. Rep.~{\bf #1} (19#3) #2}}
\def\zp#1#2#3  {{Z. Phys.~{\bf #1} (19#3) #2}}
\def\epj#1#2#3  {{Eur. Phys. J.~{\bf #1} (19#3) #2}}
\def\rmp#1#2#3  {{Rev. Mod. Phys.~{\bf #1} (19#3) #2}}
\def\JETP#1#2#3 {{Sov.\ Phys.\ JETP~{\bf #1} (19#3) #2}}
\def\sj#1#2#3 {{Sov.\ J.\ Nucl.\ Phys.~{\bf #1} (19#3) #2}}
\def\hepph  #1 {{\tt hep-ph/#1}}

\begin{document}

\begin{flushright}
YITP-99-69\\
NIKHEF/2000-002 \\
\today\\
\end{flushright}

\vspace*{30mm}

\begin{center}
{\LARGE Higher-Order QCD Corrections

\medskip

in Prompt Photon Production}
\par\vspace*{20mm}\par
{\large Eric Laenen$^a$,
George Sterman$^b$ and Werner Vogelsang$^b$}

\bigskip

{\em $^a$NIKHEF Theory Group, Kruislaan 409\\ 1098 SJ Amsterdam, The
Netherlands}

\bigskip

{\em $^b$C.N.\ Yang Institute for Theoretical Physics,
SUNY Stony Brook\\
Stony Brook, New York 11794 -- 3840, U.S.A.}
\end{center}
\vspace*{15mm}

\begin{abstract}
We exhibit a method for simultaneously treating recoil and
threshold corrections in single-photon inclusive
cross sections, working within the formalism
of collinear factorization.
This approach conserves both the
energy and transverse momentum of resummed radiation.
At moderate $p_T$, we find the potential for substantial enhancements
from higher-order perturbative and power-law nonperturbative
corrections.
\end{abstract}

Collinear
factorization for short distance, inclusive cross sections
is an important tool of particle physics.
Next-to-leading order (NLO) corrections in the strong coupling $\as$ are
known for a wide, and still-growing, set of reactions.
NLO cross sections, however, are not uniformly successful in describing
hard-scattering data, which has suggested the necessity of examining
yet higher orders in $\as$.
Among the widely-discussed examples is prompt, or direct, photon
production \cite{dphoton,e706}.

In this letter, we describe a new approach for studying higher-order
corrections in the context of collinear
factorization for the prompt photon cross section,
\be
p_T^3 {d \sigma_{AB\to \gamma X}(x_T^2) \over dp_T }
=
\sum_{ab}
\int dx_a\; \phi_{a/A}(x_a,\mu)\,
\int dx_b\; \phi_{b/B}(x_b,\mu)\;
p_T^3{d\hat \sigma_{ab\to \gamma X}\left(\hat
x_T^2,\mu\right)
\over dp_T}\; ,
\label{dgamptcofact}
\ee
where
$d\hat\sigma_{ab\to\gamma X}/dp_T$ is the hard-scattering
function at fixed $p_T$.
Hadronic and partonic scaling variables are
$x_T^2 \equiv { 4p_T^2/S}$ and  $\hat x_T^2 \equiv { 4p_T^2/\hat s}$,
respectively, with $\hat s=x_ax_bS$ the partonic center-of-mass
energy squared.
As $\hat s$ approaches its
minimum value at $\hat x_T^2=1$, the phase space available for
gluon bremsstrahlung vanishes, which results in large corrections to
$d\hat\sigma/dp_T$
at all orders.  Threshold resummation
\cite{dyresum,1pIresum,thrphen}
organizes this singular, but integrable, behavior of
$d\hat\sigma/dp_T$.

Another source of
higher-order corrections is the recoil of the
observed particle against unobserved radiation.  Thus, for Eq.\
(\ref{dgamptcofact}),
beginning with NLO in $d\hat\sigma/dp_T$,
a soft gluon may be radiated before the hard scattering, say a QCD-Compton
process $g\, q\to \gamma\, q$.  The outgoing $\gamma\, q$ pair recoils against
the soft gluon, and, as a result,
the Compton process may be softer than would be the case without
initial-state radiation.

Valuable insights have emerged in studies of
direct photon production in which the partons initiating the hard scattering
are described by generalized parton distributions
\cite{kteff1,LiLai}.  Such distributions include
transverse momentum that is partly intrinsic
and partly perturbative, whether formulated in
impact parameter ($b$) space \cite{yuancollab}
or in transverse momentum space \cite{evst,fnr}.
Nevertheless, it is difficult to state confidently whether intrinsic
transverse momentum is required by the data, which themselves allow varied
interpretations \cite{afgkpw}.  Li \cite{Liunified} showed first
how to develop a joint resummation in both leading threshold and
transverse momentum logarithms in parton distributions.

Our approach remains within the formalism
of collinear factorization.   Contributions
to the hard scattering function associated with threshold resummation
are redistributed over soft
gluon transverse momenta, simultaneously conserving
energy and transverse momentum.  Accounting for
recoil leads to an additional enhancement.

To compute higher-order recoil effects in  Eq.\ (\ref{dgamptcofact}),
we consider the partonic cross section \cite{1pIresum} for $a\, b 
\to\gamma\, c$,
and fix the center-of-mass (c.m.) rapidity, $\eta$ of the photon.
Near threshold, the overall process consists of a hard $2\to 2$ subprocess,
along with soft radiation, which can be factorized.  To leading power,
$1/Q_T^2$,
this soft  radiation does not change the
flavor of the initial-state partons,
and the hard-scattering subprocess
recoils from soft radiation with a transverse momentum ${\bf Q}_T$.
For fixed ${\bf p}_T$, the transverse momentum of the
photon relative to the c.m.\ of the hard scattering is ${\bf p}_T-{\bf Q}_T/2$.
The c.m.\ rapidity
is thus related to ${\bf Q}_T$ by
\be
{1\over \cosh^2 \eta} = {4|{\bf p}_T-{\bf Q}_T/2|^2\over \tilde s }
\equiv { 4|{\bf p}'_T|^2 \over \tilde s} \equiv \tilde x_T^2 \, ,
\label{tildexdef}
\ee
where $\tilde s\le S$ is the c.m.\ energy
squared of the $2\rightarrow 2$ hard subprocess.  The relative
transverse momentum, ${\bf p}_T'$ thus sets the minimum
value of $\tilde s$, and the scaling variable
for the subprocess is $\tilde x_T^2$.
This kinematic linkage between transverse momentum
and partonic energy drives the quantitative effects of recoil.
We must limit the size of $Q_T$ in this
analysis by some matching scale, to avoid going outside the
region where the singularities in $Q_T$ dominate.  For
small $Q_T$, however, we can jointly resum large corrections
to $d \sigma_{ab\to \gamma c}/ d^2{\bf Q}_T\, dp_T$, in
logarithms of $1-\hat x_T^2$ and $Q_T$.
The latter cancel in the hard-scattering functions.
Like threshold corrections, however, they may leave finite remainders.

In summary, the resummed inclusive cross section for the leading contributions
to prompt photon production is of the form:
\be
p_T^3{ d \sigma_{ab\to \gamma c}^{({\rm resum})}(\bar\mu) \over dp_T}
\sim
      \int d^2{\bf Q}_T\; p_T^3 {d \sigma_{ab\to \gamma c}^{({\rm resum})} \over
d^2{\bf Q}_T\, dp_T}\; \Theta\left(\bar{\mu}-Q_T\right)\, ,
\label{formal2}
\ee
where $\bar\mu$ is a matching scale.  On the right-hand side,
the direction of ${\bf  p}_T$ may be chosen arbitrarily, because of
the azimuthal symmetry of the overall process.
The inclusive partonic cross section
$p_T^3{ d \sigma_{ab\to \gamma c}^{({\rm resum})} / dp_T}$
in Eq.\ (\ref{formal2}) determines the resummed hard-scattering
function $p_T^3 d\hat\sigma_{ab\to\gamma c}^{({\rm resum})}/dp_T$
in Eq.\ (\ref{dgamptcofact}), after collinear factorization.

The arguments leading to a jointly resummed formula for
${d \sigma_{ab\to \gamma c}^{({\rm resum})} /d^2{\bf Q}_T\, dp_T}$
are similar to
those for transverse momentum resummation in Drell-Yan production.
The possibility of joint resummation for singular behavior in $Q_T$
and $1-\hat x_T^2$ is ensured by the factorization properties of
the partonic cross section near threshold \cite{cttwcls}, which
we assume here. The resummed cross section that results from
these considerations \cite{LSWprep} is
\ba
{p_T^3 d \sigma^{({\rm resum})}_{AB\to \gamma X} \over dp_T}
&=& \sum_{ij} \frac{p_T^4}{8 \pi S^2} \int_{\cal C} {dN \over 2 \pi i}\;
\tilde{\phi}_{i/A}(N,\mu) \tilde{\phi}_{j/B}(N,\mu)\;
\int_0^1 d\tilde x^2_T \left(\tilde x^2_T \right)^N
{|M_{ij}(\tilde x^2_T)|^2\over \sqrt{1-\tilde{x}_T^2}}
\nonumber \\
&& \hspace{-5mm} \times
\int {d^2 {\bf Q}_T \over (2\pi)^2}\;
\Theta\left(\bar{\mu}-Q_T\right)
\left( \frac{S}{4 {\bf p}_T'{}^2} \right)^{N+1}\;
P_{ij}\left( N,{\bf Q}_T,\frac{2 p_T}{\tilde x_T},\mu \right)\, ,
\label{1pIresumE}
\ea
in terms of moments of the physical, hadronic parton distributions
and of the squared $2\to 2$ amplitude $|M_{ij}|^2$, with ${\bf p}_T'$
defined in (\ref{tildexdef}).
The function $P_{ij}(N,{\bf Q}_T)$
is a ``profile" of ${\bf Q}_T$-dependence for fixed $N$.
Recoil is incorporated in Eq.\ (\ref{1pIresumE})
through the interplay of the profile function
and the factor $(S/4{\bf p}_T'{}^2)^{N+1}$.
Singular ${\bf Q}_T$ behavior is most easily organized in impact parameter
space, where logarithms of both the transform variables $b$ and $N$
exponentiate.
We may thus write the profile functions in Eq.\ (\ref{1pIresumE}) as
\be
P_{ij}\left( N,{\bf Q}_T,Q,\mu \right)
=
\int d^2 {\bf b} \,
{\rm e}^{-i {\bf b} \cdot {\bf Q}_T} \,
\exp\left[E_{ij\to \gamma k}\left( N,b,Q,\mu \right)\right]\, ,
\label{Pdef}
\ee
where azimuthal symmetry insures that $E$ is a function of
$b=|{\bf b}|$ only.
Here and below, $Q\equiv 2p_T/\tilde x_T$ denotes the hard
scale in the exponent.
The parameter $\mu$ represents both the
renormalization scale and the factorization scale, whose
explicit dependence we denote by $\mu_f$ below.
Eq.\ (\ref{1pIresumE}) reverts to
a threshold-resummed
prompt photon cross section when recoil is neglected, by setting
$b$ to zero in the exponents $E$ of~(\ref{Pdef}).

We now construct the exponents $E_{ij\to \gamma k}$, to NLL
in both $b$ and $N$.
As already noted by Li and Lai \cite{LiLai}, logarithmic recoil
corrections (logarithms of $b$) are generated
entirely through initial-state radiation.
Final state interactions, however, do produce logarithms of $N$.
We may thus conveniently
split each exponent into initial- and final-state parts:
\be
E_{ij\to \gamma k}(N,b,Q,\mu)
=
E_{ij}^{\rm IS}(N,b,Q,\mu) + E_{ijk}^{\rm FS}(N,Q,\mu)\, .
\label{splitE}
\ee
In general, resummation of perturbative logarithms leads
to nonperturbative power corrections in both $b$ and $N/Q$.
Because our primary interest is in transverse momentum
distributions, however, we incorporate only a term proportional
to $b^2$ in $E^{\rm IS}$ \cite{yuancollab},
associated with the region of strong
coupling \cite{ks95}.
The initial-state exponents $E_{ij}^{\rm IS}$ will therefore
be a sum of perturbative and nonperturbative contributions.
The final state exponent, $E^{\rm FS}$ in Eq.\ (\ref{splitE}) is
essentially the
same as for pure threshold resummation  \cite{1pIresum}.

Rather than giving a formal derivation \cite{LSWprep} of the NLL
perturbative exponent, we motivate our expression by requiring it to
reproduce both threshold and $k_T$ resummations in the appropriate limits.
The following expression for the initial state exponent, including
a nonperturbative term, gives all leading and next-to-leading logarithms in
$b$ and $N$:
\be
E_{ij}^{\rm IS}(N,b,Q,\mu)
=
\int_{Q\chi^{-1}(N,b)}^{\mu}
{d\mu'\over \mu'}\;
\left[\, A_i(\as(\mu'{}^2))+ A_j(\as(\mu'{}^2))\, \right]\, \;
2\ln{\bar N\mu'\over Q}
- b^2\, F_{ij}(N,Q)
\, ,
\label{Elog}
\ee
where $\bar N\equiv Ne^{\gamma_E}$ with $\gamma_E$ the Euler constant.
At this point, the coefficient $F_{ij}(N,Q)$ is arbitrary.
The function $A_a(\as)\equiv\sum_n(\as/\pi)^nA_a^{(n)}$ is, as
usual \cite{KT}, $A_a^{(1)} = C_a$ and
$$A_a^{(2)} = \frac{1}{2} C_a K \equiv \frac{1}{2} C_a \left[ C_A \left(
\frac{67}{18} - \frac{\pi^2}{6} \right) -\frac{10}{9} T_R N_f \right]\, ,$$
with $C_q=C_F$, $C_g=C_A$.

The perturbative exponent in Eq.\ (\ref{Elog}) is similar to
initial-state contributions in
threshold  resummation \cite{1pIresum,thrphen}, but with a new lower limit for
the integral over the coupling scale $\mu'$,
which we denote $Q\chi^{-1}(N,b)$.
The following simple choice of $\chi$ gives all NLL logarithms:
\be
\chi(N,b)
=
     \bar N + bQ/c_1 \, ,
\label{chidef}
\ee
where $c_1=2 {\rm e}^{-\gamma_E}$.
This doubly-resummed NLL exponent has the important
property of respecting momentum conservation, not only for transverse
components as in $k_T$-resummation, but for energy as well.

The exponent in (\ref{Elog}) is a function of the logarithmic variables,
\begin{eqnarray}
\lambda &=& b_0 \alpha_s (\mu^2)  \ln \bar{N} 
\nonumber \\
\beta &=& b_0 \alpha_s (\mu^2)
\ln \left(\chi(N,b) \right) \, ,
\label{vars}
\ea
where
$b_0 = (12\pi)^{-1}(11 C_A - 4 T_R N_f)$.
The initial state
contributions are given in these terms by
\be
E_{ij}^{\rm IS}(N,b,Q,\mu) = \sum_{a=i,j}\ \left[
\frac{1}{\alpha_s (\mu^2)} h^{(0)}_a (\lambda,\beta) +
h^{(1)}_a (\lambda,\beta,Q,\mu,\mu_f)  \right]\, ,
\label{Hish}
\ee
where, adopting the notation of Ref.\ \cite{thrphen,fnr},
\begin{eqnarray}
h_a^{(0)} (\lambda,\beta) &=& \frac{A_a^{(1)}}{2\pi b_0^2}
\left[ 2 \beta + (1 - 2 \lambda) \ln(1-2 \beta) \right]\, , \\
h_a^{(1)} (\lambda,\beta,Q,\mu,\mu_f) &=&
\frac{A_a^{(1)} b_1}{2\pi b_0^3} \left[ \frac{1}{2} \ln^2 (1-2 \beta) +
\frac{1 - 2 \lambda}{1-2\beta} (2 \beta + \ln(1-2 \beta) ) \right] \nonumber \\
&\ & \hspace{-42mm} + \frac{1}{2\pi b_0} \left( - \frac{A_a^{(2)}}{\pi b_0} +
A_a^{(1)} \ln \left( \frac{Q^2}{\mu^2} \right) \right)
\left[ 2 \beta \frac{1 - 2 \lambda}{1-2\beta}+ \ln(1-2 \beta) \right]
- \frac{A_a^{(1)}}{\pi b_0} \lambda \ln \left( \frac{Q^2}{\mu_f^2}
\right),
\label{hsubadef}
\end{eqnarray}
with $b_1=(17 C_A^2-10 C_A T_R N_F -6 C_F T_R N_F)/24 \pi^2$.

Given the above, we are in a position to illustrate the role of recoil,
using Eq.\ (\ref{1pIresumE}),
through the evaluation of the $N$ and $Q_T$
integrals, with
profile functions at fixed $N$ defined by Eqs.\ (\ref{Pdef})-(\ref{hsubadef}).
As in threshold resummation, it is necessary to define the integrals in a
manner that avoids the (``Landau") singularities in the running coupling
of Eq.\ (\ref{Elog}) at large $N$ and/or $b$.
For pure threshold resummation, a number of methods have
been employed to define the $N$ integral in (\ref{1pIresumE}).  We will use
a ``minimal" approach \cite{thrphen,cmnt},  defining the
contour ${\cal C}$ in (\ref{1pIresumE})
to extend to infinity through the negative real half-plane.
For the $b$ integral in the profile
function, Eq.\ (\ref{Pdef}), we have developed a new integration
technique.

To perform the $b$ integral, we rewrite Eq.\ (\ref{Pdef}) as
\be
P_{ij}\left( N,{\bf Q}_T,Q,\mu \right)
    =
\pi\, \int_0^\infty db\, b\, \left[\, h_1(bQ_T,v) + h_2(bQ_T,v)\, \right]\,
{\rm e}^{E_{ij\to \gamma k}\left( N,b,Q,\mu \right)}\, .
\label{J0split}
\ee
Here we introduce two auxiliary functions $h_{1,2}(z,v)$, related
to Hankel functions and defined in terms of an arbitrary real,
positive parameter $v$
by integrals in the complex $\theta$-plane:
\ba
h_1(z,v)
&\equiv& - {1\over\pi}\
\int_{-iv\pi}^{-\pi+iv\pi}\, d\theta\, {\rm e}^{-iz\, \sin\theta}
\nonumber\\
h_2(z,v)
&\equiv& - {1\over\pi}\
\int^{-iv\pi}_{\pi+iv\pi}\, d\theta\, {\rm e}^{-iz\, \sin\theta}\, .
\label{hdefs}
\ea
The $h_{1,2}$ become the usual Hankel functions $H_{1,2}(z)$ in the
limit $v\rightarrow \infty$.  They are finite for any finite
values of $z$ and $v$.  Their sum is always:
$h_1(z,v)+h_2(z,v)=2J_0(z)$, independent
of $v$. The utility of the $h$-functions is that they distinguish
positive and negative phases in Eq.\ (\ref{J0split}),
making it possible to treat the $b$ integral of
the profile function as the sum of two contours,
one for each $h_i$.  These contours avoid
the Landau pole by a deformation into either the upper half-plane
($h_1$), or the lower half-plane ($h_2$).  Such a definition of
the integral is completely equivalent to the original form,
Eq.\ (\ref{Pdef}) when the exponent is evaluated to finite
order in perturbation theory.  It defines the resummed integral
``minimally" \cite{cmnt},
without an explicit cutoff.

Proton-nucleon cross sections computed in this fashion are illustrated in
Fig.\ \ref{profiles}.  Here we show, for several values of photon $p_T$,
$d\sigma^{(\rm resum)}_{{\rm pN}\to\gamma X}/dQ_Tdp_T$, the distribution of
the cross section in the recoil momentum $Q_T$.
The kinematics are those of the E706 experiment~\cite{e706}.  Since this is
a ``demonstration"
calculation, we pick a nominal value of $F_{ij}(N,Q)=0.5$ GeV$^2$ for the
Gaussian coefficient in Eq.\ (\ref{Elog}), independent of parton type.
We leave for future work a more realistic determination of this coefficient,
including its $Q$-dependence.  The parton distributions
are those of Ref.\ \cite{GRV}, and we treat NLO $N$-independent 
(``hard virtual'') terms
as in \cite{thrphen}.  For simplicity, we approximate $Q$ in Eq.\ (\ref{Pdef})
by $2p_T$.  Finally, we set $v=1$ in Eq.\ (\ref{hdefs}).

The  dashed lines are $d\sigma^{(\rm resum)}_{{\rm pN}\to\gamma
X}/dQ_Tdp_T$ from
Eq.\ (\ref{1pIresumE}), but with recoil neglected by fixing $S/4{\bf p}_T'{}^2$
at $S/4 {\bf p}_T^2$.
The dashed lines thus show how each $Q_T$ contributes to
threshold enhancement.
The profiles are subtracted in their $b$-dependent one-loop
corrections, which produce the peaks and dips at low $Q_T$ for
the smaller values of $p_T$. Every curve shows a peak (near 2 GeV)
associated with resummation, and falls off as a power with increasing $Q_T$.
Integrating the dashed lines reproduces the threshold-resummed cross
sections of Ref.\ \cite{thrphen}, after a subtraction at order $\as$ to
recover exact NLO.

The solid lines of Fig.\ \ref{profiles} show the $N$-integrated
distributions in $Q_T$, \linebreak[4]
$d\sigma^{(\rm resum)}_{{\rm pN}\to\gamma X}/dQ_Tdp_T$, now
found by including the true recoil factor $(S/4{\bf p}_T'{}^2)^{N+1}$ in Eq.\
(\ref{1pIresumE}).  These curves
therefore describe the fully resummed cross section.
The resulting enhancement is clearly substantial.  For small $p_T$, the
enhancement simply grows with $Q_T$ (it must diverge at $Q_T=2p_T$),
while for $p_T$ above 5 GeV it has a dip at about $Q_T = 5$ GeV,
which becomes more pronounced
as $p_T$ increases, but which never really gets to zero.  This behavior
makes it problematic to identify a unique matching scale $\bar\mu$ in
(\ref{1pIresumE})
that gives a stable prediction for the cross section.

\begin{figure}[tbh]
\begin{center}
\hspace*{-7mm}
\vspace*{3mm}
\epsfig{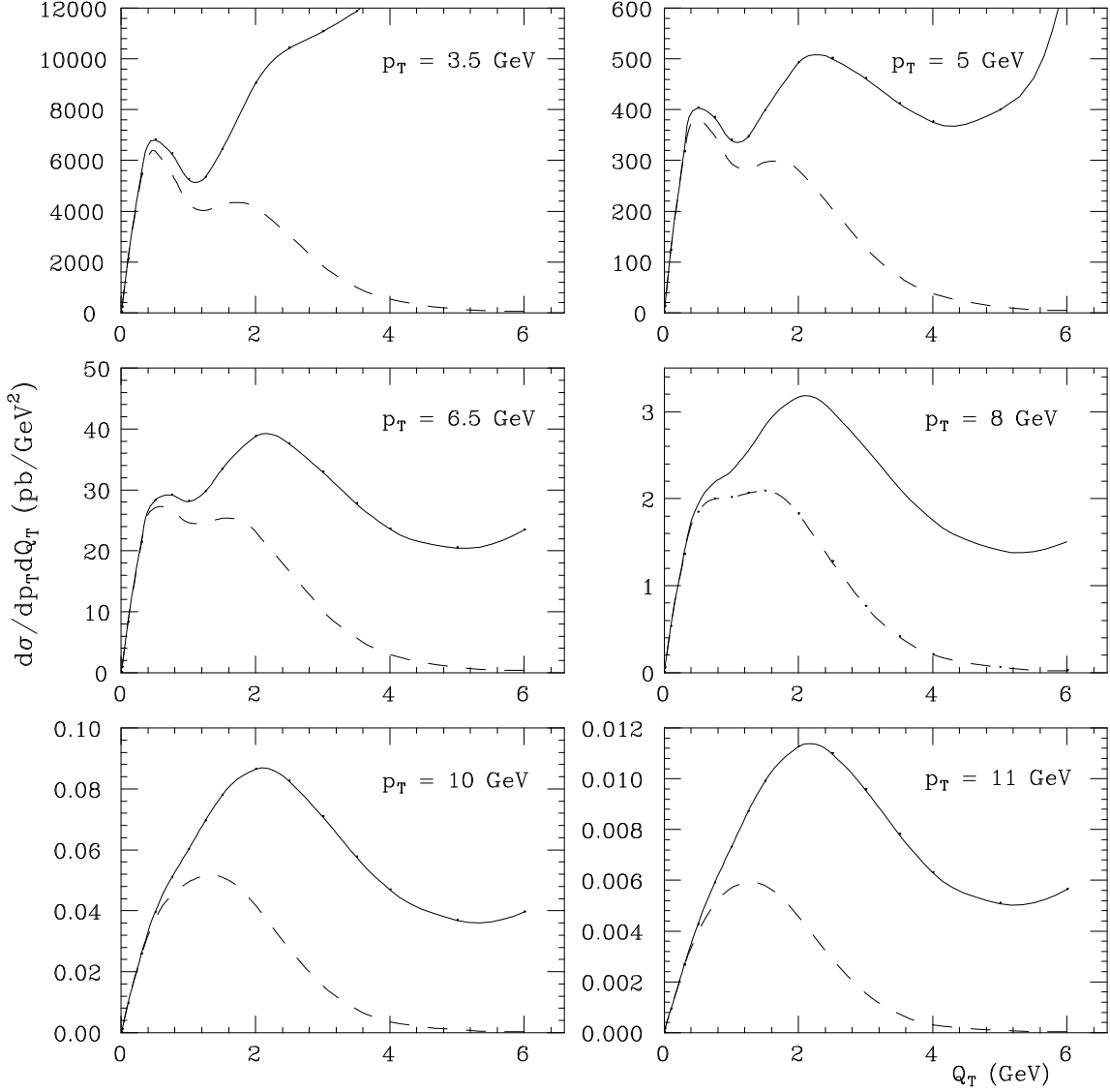}
\end{center}
\caption{The prompt photon cross section
$d\sigma_{{\rm pN}\to\gamma X}/dQ_Tdp_T$ at {\protect $\sqrt{s}=31.5$} GeV,
as a function of $Q_T$ for various values of photon $p_T$. Dashed lines
are computed without recoil ({\protect
${\bf p}_T'={\bf p}_T$} in~(\ref{1pIresumE})), solid lines are with recoil.}
\label{profiles}
\end{figure}

This reservation notwithstanding, and reemphasizing that our
calculation is
primarily an illustration, rather than a quantitative prediction, we
plot in Fig.\ \ref{crossec} the resummed
cross section for $p_T\ge 3.5$ GeV, with the
choice $\bar\mu=5$ GeV for the calculation
of Fig.\ \ref{profiles}, using the approximate procedure of Ref.\
\cite{thrphen}
to convert $d\sigma_{{\rm pN}\to\gamma X}/dp_T$ to a cross
section integrated
over a finite rapidity interval. We include an NLO photon
fragmentation component in the cross section, again calculated as
in~\cite{thrphen}. To show the size of the
enhancement that recoil can produce, as well as its potential
phenomenological impact, we also exhibit the pure threshold
resummed cross section and the E706
direct photon data in this range \cite{e706}.  For the theoretical
curves, we choose $\mu=p_T$; both resummed curves have sharply reduced
factorization scale dependence compared to NLO.
Evidently, recoil can be phenomenologically relevant.

\begin{figure}[tbh]
\begin{center}
\hspace*{-7mm}
\vspace*{3mm}
\epsfig{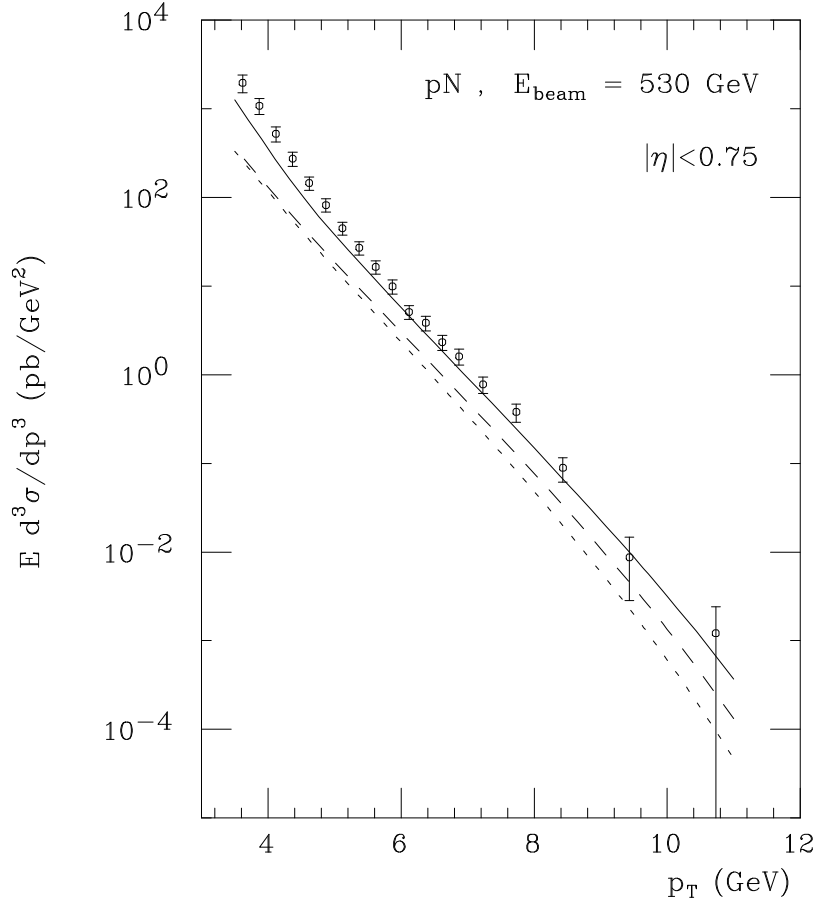}
\end{center}
\caption{Prompt photon cross section $Ed^3 \sigma_{{\rm pN}\to\gamma X}/dp^3$
for pN collisions at {\protect $\sqrt{s}=31.5$ GeV}. The dotted line
represents the full NLO calculation, while the dashed and solid lines
respectively incorporate pure threshold
resummation~{\protect \cite{thrphen}} and the joint resummation described
in this paper.  Data have been taken from~{\protect \cite{e706}}.}
\label{crossec}
\end{figure}

It will take some time to explore
this resummation formalism, including the implementation
of practical nonperturbative estimates and of matching procedures.
Nevertheless, we hope that this formulation
of recoil effects at higher orders, in
the language of collinear factorization,
is a step toward clarifying what
has been a thorny issue in the application of perturbative QCD.
We also anticipate that the joining of threshold and transverse momentum
resummation \cite{Liunified}, along with the minimal evaluation of
the profile function described above, will have useful applications
to Drell-Yan and related cross sections.

\subsection*{Acknowledgements}  We thank M.\ Fontannaz, J.\ Huston, 
N.\ Kidonakis
J.\ Owens and M.\ Zielinski for pertinent discussions.
The work of G.S.\ and W.V.\ was supported in part by the National 
Science Foundation,
grant PHY9722101. The work of E.L.\ is part of the research program of the
Foundation for Fundamental Research of Matter (FOM) and
the National Organization for Scientific Research (NWO).

\end{document}